\documentclass[runningheads]{llncs}
\usepackage[T1]{fontenc}
\usepackage{graphicx}
\usepackage{amssymb}
\usepackage{hyperref}
\usepackage{color}

\begin{document}
\title{Eddeep: Fast eddy-current distortion correction for diffusion MRI with deep learning}
\titlerunning{Eddeep: Fast eddy-current distortion correction for diffusion MRI}
%

\author{Antoine Legouhy\inst{1,2}, Ross Callaghan\inst{2}, Whitney Stee\inst{3,4}, Philippe Peigneux\inst{3,4}, Hojjat Azadbakht\inst{2}, Hui Zhang\inst{1}}
\authorrunning{A. Legouhy et al.}
%
\institute{Centre for Medical Image Computing \& Department of Computer Science, University College London, London, UK
\and AINOSTICS ltd., Manchester, UK
\and UR2NF-Neuropsychology and Functional Neuroimaging Research Unit affiliated at CRCN – Centre for Research in Cognition and Neurosciences and UNI - ULB Neuroscience Institute, Université Libre de Bruxelles (ULB), Brussels, Belgium
\and GIGA - Cyclotron Research Centre - In Vivo Imaging, University of Liège (ULiège), Liège, Belgium}
\maketitle              
\begin{abstract}

Modern diffusion MRI sequences commonly acquire a large number of volumes with diffusion sensitization gradients of differing strengths or directions. Such sequences rely on echo-planar imaging (EPI) to achieve reasonable scan duration. However, EPI is vulnerable to off-resonance effects, leading to tissue susceptibility and eddy-current induced distortions. The latter is particularly problematic because it causes misalignment between volumes, disrupting downstream modelling and analysis. The essential correction of eddy distortions is typically done post-acquisition, with image registration. However, this is non-trivial because correspondence between volumes can be severely disrupted due to volume-specific signal attenuations induced by varying directions and strengths of the applied gradients. This challenge has been successfully addressed by the popular FSL~Eddy tool but at considerable computational cost. We propose an alternative approach, leveraging recent advances in image processing enabled by deep learning (DL). It consists of two convolutional neural networks: 1) An image translator to restore correspondence between images; 2) A registration model to align the translated images. Results demonstrate comparable distortion estimates to FSL~Eddy, while requiring only modest training sample sizes. This work, to the best of our knowledge, is the first to tackle this problem with deep learning. Together with recently developed DL-based susceptibility correction techniques, they pave the way for real-time preprocessing of diffusion MRI, facilitating its wider uptake in the clinic.

\end{abstract}

\section{Introduction}

In magnetic resonance imaging (MRI), diffusion-weighted imaging (DWI) is designed to indirectly uncover the intricate microstructure of biological tissues through the study of the movement of their water molecules. It has various clinical applications including stroke diagnostic~\cite{fiebach2002} or tumor characterisation~\cite{maier2010}. 
DWI requires the acquisition of a large number of volumes with different diffusion sensitization gradients, more than a hundred for advanced diffusion models. To achieve acceptable scan duration, spin-echo single-shot echo-planar imaging (EPI) sequences are used~\cite{turner1990,skaare2010}. However, this technique is vulnerable to off-resonance effects leading to tissue susceptibility and eddy-current induced distortions. Eddy currents are a result of rapid switching of magnetic field gradients. They induce additional unwanted spatially-varying magnetic fields that will interfere with the spatial localization, leading to geometrical distortions~\cite{chang1992,jezzard1998}. As the nature and extent of the distortions depend on the strength and direction of the applied diffusion gradient, this causes misalignment between the volumes of a single DW acquisition, thus disrupting downstream modelling and analysis. 

The essential correction of eddy distortions is typically done post-acquisition, with image registration. This task is non-trivial because correspondences between volumes are severely disrupted by the volume-specific signal attenuations induced by varying directions and strengths of the applied gradients. In particular, the outer cerebro-spinal fluid (CSF) is a vanishing boundary beyond very low b-values, very different attenuation patterns appear in areas of anisotropic diffusivity like white matter, and the signal-to-noise ratio (SNR) decreases as the b-value increases. As a result, even intensity similarity metrics designed to handle large differences in contrast, like mutual information, do not behave well in general. Therefore, although free of eddy distortions, a $b=0$ image cannot directly be used as target target for registration. 

Strategies have been developed to address this challenge. In FSL~Eddy~\cite{fsleddy}, arguably the most popular tool for correcting eddy-current distortion, instead of targeting a $b=0$, each volume is iteratively registered to a predicted, non-distorted, version of itself through a Gaussian process~\cite{gproc}. 
In Tortoise~\cite{tortoise}, the correction strategy is adapted to the b-value of each volume. But except for the low b-value case, where volumes are corrected by registering with mutual information as the similarity metric, the essential idea is also to predict a non-distorted version of each volume, albeit with a different, b-value-dependent, approach compared to FSL~Eddy. 
These tools achieve good correction but they are computationally intensive as they rely on traditional registration which involves restarting a new optimisation from scratch each time.

Recently, new approaches to geometric distortion corrections have been developed, taking advantage of major advances in image processing enabled by deep-learning (DL). In particular DL-based image registration methods~\cite{devos2019,voxelmorph1} demonstrated state-of-the-art alignment accuracy, doing so in a split second at inference. These have already successfully been applied to the correction of susceptibility distortions~\cite{duong2020,anon1}, but not, to the best of our knowledge, to eddy distortions, likely because of the difficulty to handle disrupted correspondences. We hypothesize that this difficulty may be addressed by leveraging similar advances in image translation enabled by DL where conditional GAN (cGAN)~\cite{mirza2014} architectures like pix2pix~\cite{isola2017} have been used in medical imaging for T1-weighted to $b=0$ image translation in~\cite{schilling2019}.

In this paper, we propose a novel approach for the correction of eddy-current distortions that leverages those deep-learning innovations. It is composed of two models: 1) A 3D pix2pix image translator, a paired and supervised model, to restore correspondences between volumes. Regardless of its b-value and gradient direction, each volume is translated to a version that corresponds to a target b-value and is orientation-averaged. 2) An unsupervised registration model to align the translated images. Together, they allow us to estimate and correct for eddy distortion and head motion. 

We trained and evaluated five models with various training sizes and augmentation configurations. Results demonstrate comparable distortion correction to FSL~Eddy, while requiring only modest training sample sizes. Correction at inference is orders of magnitude faster than traditional techniques. 

\section{Background}
\subsection{Eddy distortion model}
\label{distomodel}

The theory behind distortions induced by unwanted, additional fields to the $B_0$ is detailed in~\cite{chang1992}. The particular case of eddy distortions for DWI have been studied in~\cite{jezzard1998}. 
It emerges that eddy distortions can reach an amplitude of several millimeters along the phase-encoding direction (PED) but are negligible along the remaining directions; only DW images are affected (not $b=0$) and the distortions varies with respect to the diffusion gradient strength; and that the induced shift, to the first order approximation, may be expressed as a linear combination of the original coordinates. However, it has empirically been shown in~\cite{fsleddy,rohde2004} that this linear approximation is not always adequate and that additional higher-order terms may be required. A quadratic model has been proposed as a good trade-off between increased flexibility and parsimony.

Let $x=(x_1,x_2,x_3)^T$ be the original coordinates of a voxel and $y=(y_1,y_2,y_3)^T$ be the distorted counterparts. Assuming the PED along the 2nd axis ($y_1=x_1$ and $y_3=x_3$) and, the quadratic distortion model, $y_2$ can be expressed as:
\begin{equation}
y_2 = x_2 + x^TQx+Lx+t\ ,\quad\quad\textrm{ where } Q\in \mathbb{R}^{3\times 3}_{\mathrm{sym}},\ L\in \mathbb{R}^{1\times 3} \textrm{ and } t\in \mathbb{R}.\vspace{0.1cm} 
\end{equation}
This uni-directional (constrained along the PED) global transformation has 10 degrees of freedom (dof) in 3D. 

\subsection{Accounting for between-volumes head motion}
Due to the speed of EPI, within-volume head motion may be considered negligible. Between-volumes head motion can be estimated using a rigid transformation (6 dof in 3D). The order matters when composing the estimated motion matrix $R$ with the estimated eddy distortions $E$. The subject's head can move freely in space but distortions are image reconstruction artifacts that are anchored to the reference frame defined by the imaging gradients (slice-select, phase-encoding and frequency-encoding). The coordinates accounting for motion should be written: $y=E\circ Rx$. Eddy and rigid transformations are disentangled for better interpretability but their parameters are not all independent.

\section{Methods}
\begin{figure}
    \centering
    \includegraphics[width=\linewidth]{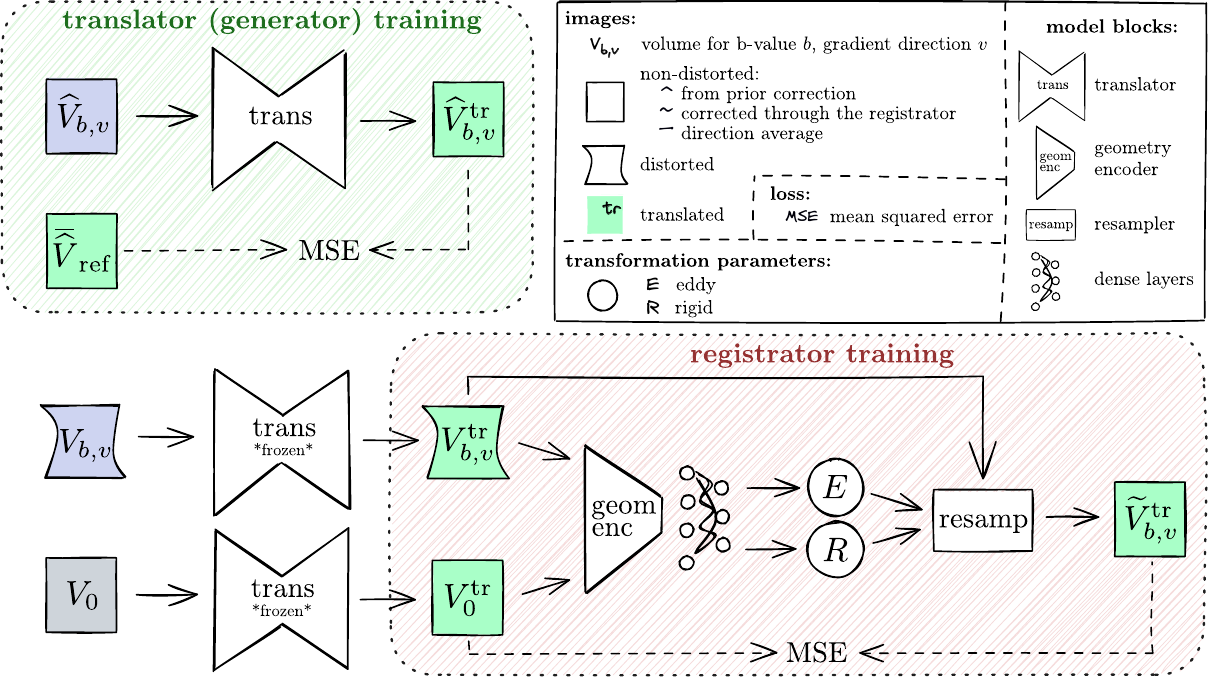}
    \caption{Overview of the generator part of the translator and the registration model during training.}
    \label{diagmodel}
\end{figure}

An overview of our approach is shown in Fig.~\ref{diagmodel}. The rest of this section describes the image translator and the image registration model used in turn for eddy distortion correction, before discussing the augmentation strategies for training. 

\subsection{Image translation}

The choice of the target for translation is guided by the following observations:
\begin{itemize}
   \item For a given b-value, assuming a uniform sampling on the sphere of the gradient directions, averaging the distortion free volumes results in an image that has no direction dependency.
   \item It is easier to remove extra material around an object (outer CSF signal at low b-value) than to generate anything that is absent.
   \item At moderately high b-value, typically between 700 and 3000, the CSF signal is fully attenuated to noise level, but the signal-to-noise ratio in tissue is still sufficiently high to preserve edges that drive alignment.
\end{itemize}
Therefore, we propose an image translator that takes a DW volume of any gradient direction and b-value (including $b=0$), and produces an homologous aligned volume corresponding to a moderate b-value $b_\mathrm{ref}$, averaged over all directions. As it is essential that the translation process does not introduce any shift, the more constrained paired cGAN architecture of pix2pix~\cite{isola2017} is preferred to the cycleGAN~\cite{zhu2017} alternative.

Given the need of pix2pix for paired training data, in the current implementation, these are produced using FSL~Eddy for simplicity. A moderately high target b-value $b_\mathrm{ref}$ is chosen among the available one. Then, for each subject, a target image for supervision is built by averaging the corrected volumes associated with $b_\mathrm{ref}$. During training, corrected data is used as input and the model learns to translate it into the target image (see Fig.~\ref{diagmodel}).
During inference, however, the translator receives distorted data as input, generating corresponding distorted outputs. Those will be the inputs of the subsequent registration model. Although eddy distortion magnitudes are significant enough to affect downstream modeling and analysis, they remain sufficiently minor for a translator trained on corrected data to effectively handle distorted one during inference. This is especially true if spatial augmentation is done on the training data.

\subsection{Image registration}
Once translated, correspondences between volumes are restored, enabling accurate intensity-based registration. In fact, contrasts are now so similar that a simple mean squared error (MSE) can be used as similarity criterion to guide a registration. Recently, the problem of intensity-based image registration has been tackled using convolutional neural networks (CNN). The case of deformable registration have been addressed in~\cite{devos2019,voxelmorph1,voxelmorph2} using a sequence of U-Net to estimate a deformation field and a spatial transformer to warp the moving image. In~\cite{devos2019}, a second architecture was also presented for affine transformation estimation where the U-Net is replaced by an encoder followed by dense layers. 

We propose a similar approach to estimate the eddy distortion and motion parameters following the unidirectional quadratic transformation depicted in Section~\ref{distomodel}. The registration model, illustrated in Fig.~\ref{diagmodel}, is a sequence of the following blocks:

\begin{enumerate}
    \item Input: A distorted, translated DW volume (moving); and a translated $b=0$ image (reference); concatenated.
    \item Geometry encoder: Input goes through a CNN encoder learning to extract features of varying spatial resolutions for localization.
    \item Dense layers: The flattened encoder output goes through a series of fully connected layers, where the nodes of the last layer are the parameters of the rigid and eddy transformations. 
    \item Resampler: A spatial transformer network~\cite{trsfnet} transform the moving image with the estimated transformation parameters. During training, the translated moving DW volume is resampled to be compared to the translated $b=0$ one. At inference, the original moving DW volume is directly resampled instead for correction.
\end{enumerate}

\begin{remark}
We freeze the weights of the translator when producing the inputs for the registration model instead of attempting some joint optimization. Indeed, without supervision, the translator could for example learn to map everything to 0 leading to a subsequent perfect 0 MSE loss for the registration model. 
\end{remark}

\subsection{Augmentation strategies}
\label{aug}
\textbf{Translator:} During training, corrected data is used, so an easy contrast augmentation is to simply create a new input from a weighted average of two volumes of the same subject. Spatial augmentation can be done by applying the same affine and deformable transformations to the input and target images.\\
\textbf{Registration model:} No contrast augmentation is needed as the inputs are translated images. For spatial augmentation, one can apply the same transformation to the inputs if it preserves parallelism with the PED. Assuming the PED being along the second axis, one can use an affine matrix $A$ such that $A_{12}=0$ and $A_{32}=0$, and a deformation field constrained along the PED. In addition, a rigid transformation can be composed on the right, for the moving volume only.

\section{Evaluations}

\subsection{Dataset}
\label{dataset}
As part of an ongoing study investigating memory learning and consolidation, we acquired paired AP and PA diffusion-weighted EPI data for 90 subjects. Each are composed of 13 $b=0$ and DW volumes for $b=650$, $1000$ and $2000$; with respectively 15, 30 and 60 gradient directions; voxel size of 2 mm isotropic.

We split our subjects into mutually exclusive sets for training, validation, and testing. We used two sizes for training: small (S) with 8 subjects for training and 4 for validation; medium (M) with 32 and 8 respectively. The testing set comprises 30 subjects. 
While it is preferable to use distinct training/validation sets for the two models, we chose to reuse half of the training/validation sets of the translator for the registration model to minimise the combined sizes (if the translator and the registration model both use size M training/validation sizes, only 48/12 subjects overall are necessary).

\subsection{Models}
We trained 4 translators with various configurations, all targeting $b_\mathrm{ref}=2000$: $T_S$, $T_S^+$, $T_M$ and $T_M^+$; where $^+$ indicates that the augmentation in Section~\ref{aug} have been used on the input with a probability 0.5, $S$ and $M$ refer to the training sample sizes in Section~\ref{dataset}. Following the same nomenclature for the training of registration model $R$, we produced the following 5 full models: $R_S\circ T_S$, $R_S^+\circ T_S^+$, $R_S^+\circ T_M^+$, $R_M\circ T_M$ and $R_M^+\circ T_M^+$.

\subsection{Implementation details}
\textbf{Translator:}                             
We used a 3D adaption of pix2pix~\cite{isola2017}. The generator is a U-Net~\cite{unet} with: (32, 64, 128, 256) encoder features, (256, 128, 64, 32, 1) decoder features; the discriminator is a patchGAN~\cite{isola2017} with: (16, 32, 64, 128, 1) encoder features; both using Leaky ReLU activation and kernel size of 3. Training parameters: learning rate of $10^{-4}$, batch size of 4, MSE loss for both the generator and patch discriminator with weights 100 and 1 respectively.\\
\textbf{Registration model:}   
The geometry encoder has (16, 32, 40, 48, 56) features, Leaky ReLU activation and kernel size of 3. The series of fully connected layers starts with the flattened encoder input, then split in 5 sets of 64 nodes, each fully connected to resp. 6, 3, 1, 3, 3 nodes for resp.: eddy quadratic, linear and translation; rigid rotation and translation parameters; using Leaky ReLU activation except for the last layer. The resampler is similar to the spatial transformer in~\cite{voxelmorph1}, adapted to handle the quadratic transformation depicted in Section~\ref{distomodel}.  Training parameters: learning rate of $10^{-5}$, batch size of 4, MSE loss.

The code is available at:~\url{https://github.com/CIG-UCL/eddeep}. It uses tools from VoxelMorph\footnote{\url{https://github.com/voxelmorph/voxelmorph}} and SimpleITK\footnote{\url{https://simpleitk.org}}.

\subsection{Results}
\begin{figure}[!h]
    \centering
    \includegraphics[width=\linewidth]{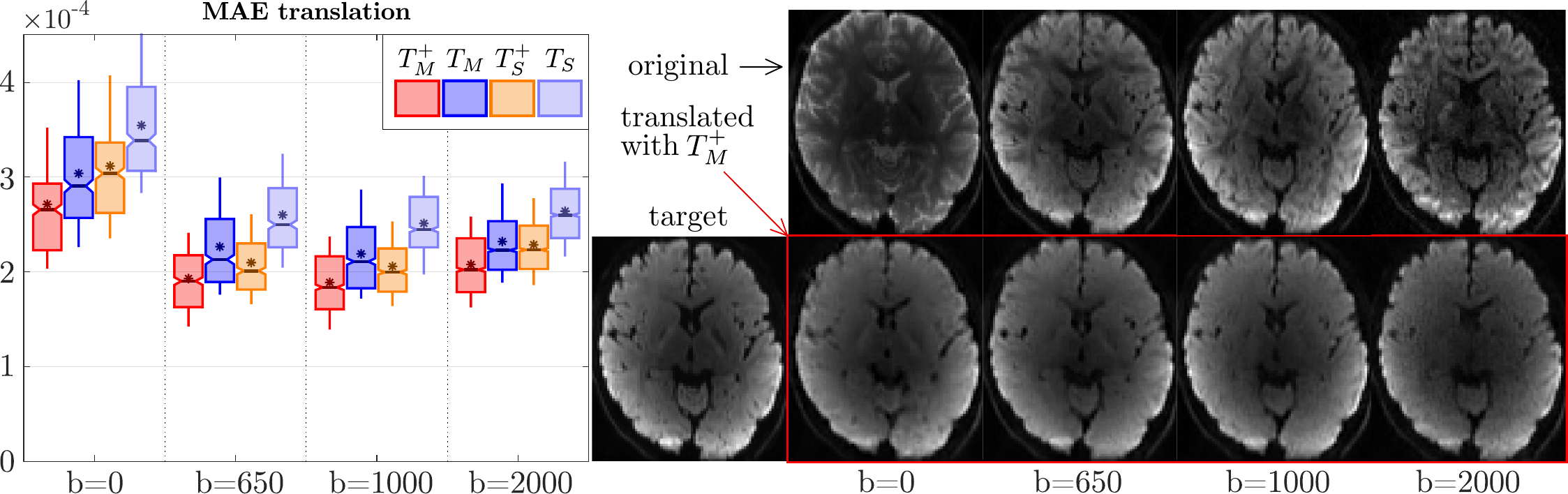}
    \caption{Translation results by b-value for the testing dataset. Boxplot: mean absolute error (MAE) between translated and target for various translators. target, original (intensity normalized) and translated images with $T_M^+$.}
    \label{transfig}
\end{figure}

\begin{figure}
    \centering
    \includegraphics[width=\linewidth]{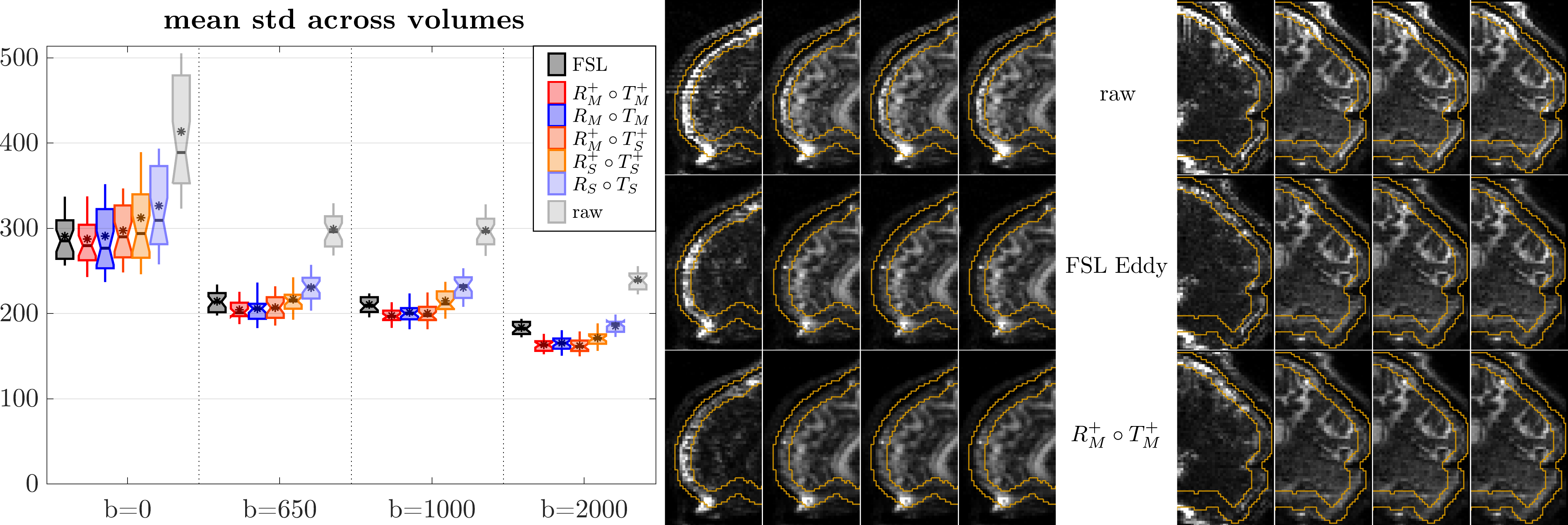}
    \caption{Correction results by b-value for the testing dataset using the standard deviation (std) across volumes. Images show the map of std for raw data (top), corrected with FSL~Eddy (middle) and through $R_M^+\circ T_M^+$ model. The outer brain mask is displayed with yellow lines. The boxplots are showing the std for uncorrected data and corrected with FSL~Eddy or our various full models.}
    \label{corrfig}
\end{figure}

Animations for qualitative assessment of the translation and distortion correction are provided as supplementary material.\\
\textbf{Translator:}  
The quality of the translators was evaluated for each b-values using MSE between the translated image and the target $b_\mathrm{ref}$ average image. Results are shown in Fig.~\ref{transfig}. As expected, for a given amount of augmentation, increasing the number of training samples leads to better results. Interestingly, with only 8 subjects and data augmentation, one can reach similar performances as with 32 subjects without augmentation.\\%
\textbf{Registration model:}   
Those models were compared against FSL~Eddy, the state-of-the-art tool for this task.
We used, for each b-value, the voxel-wise intensity standard deviation across volumes near the interface between cortical gray matter and CSF as an indicator of the amount of misalignment. This is motivated by the observation that mismatch of this interface across volumes will generate higher values of standard deviation compared to when this interface is well aligned across volumes. To quantify this, we extracted a mask of this interface (subtraction of a dilated (2 iterations) and eroded (4 iterations) brain mask from FSL BET~\cite{bet}). Results are shown in Fig.~\ref{corrfig}. Correction is better for full models where the training of the registration was done with a sample of size M. Among those, interestingly, $R_M^+\circ T_S$ does not seem to be penalized by the training sample size S of its translator. Correction with the best performing models lead to output very similar to FSL~Eddy.

\section{Discussion and conclusion}

We proposed a novel approach for the correction of eddy-current distortions, the first using with deep-learning for this task to the best of our knowledge. We use a sequence of an translator to restore correspondences between images and a registration model for distortion estimation and correction. We achieved correction on par with traditional state-of-the-art techniques. Only modest training sample sizes are necessary, especially for the translator, thus making the upfront training cost relatively low. Inference is then very rapid.
A limitations of this work is the necessity for the translator to be trained using data corrected with an external tool. However, even a translator trained with as few as 8 subjects (plus 4 for validation) leads to good correction with the subsequent registration model. It seems that, given the low amount of degrees of freedom of the distortion model, a relatively approximate translated images are sufficient.
Another limitation is the absence of ground truth distortions which limits evaluation. 
We plan to address both those shortcomings in a future work using synthetic diffusion data training input and for evaluation, in the same vein as in~\cite{graham2016} but including higher order distortion models.
Together with recently developed DL-based susceptibility correction techniques, this work paves the way for real-time preprocessing of diffusion MRI, facilitating its wider uptake in the clinic.

\begin{credits}
\subsubsection{\ackname}

This work is funded by the Medical Research Council (MRC): (MR/T046473/1; AL, HZ), Innovate UK (No. 10036158; HA, AL, HZ), Le Fonds de la Recherche Scientifique (FRS-FNRS) and the Research Foundation Flanders (FWO) as part of Excellence Of Science (EOS) project Memodyn (No. 30446199; AL, HZ, PP) and FRS-FNRS (WS).

\subsubsection{\discintname}
The authors have no competing interests to declare that are relevant to the content of this article.
\end{credits}

%
%
%
%

\end{document}